\documentclass[12pt]{book}

\usepackage{amsmath}
\usepackage[dvips]{graphicx,color}
\usepackage{makeidx,physics,cosmology}

\makeauthorindex
\makeindex

\BookTitle{Frontier in Astroparticle Physics and Cosmology}
\CopyRight{\copyright 2004 by Universal Academy Press, Inc.}

\begin{document}

\BookTitle{\itshape Frontier in Astroparticle Physics and Cosmology}
\CopyRight{\copyright 2004 by Universal Academy Press, Inc.}
\pagenumbering{arabic}

\chapter{Extracting energy from black holes: Short-GRBs, Long-GRBs and GRB afterglows}

\author{%
Remo RUFFINI\\
{\it ICRA --- International Center for Relativistic Astrophysics and Dipartimento di Fisica, Universit\`a ``La Sapienza'', Piazzale Aldo Moro 5, I-00185 Roma, Italy}\\
}
%
%
\AuthorContents{R.\ Ruffini} 

\AuthorIndex{Ruffini}{R.} 

\section*{Abstract}

The extractable energy from a black hole, as origin of the Gamma-Ray Burst (GRB) phenomenon is reviewed.

\section{Introduction}

\begin{figure}[t]
\begin{center}
\includegraphics[width=8cm,clip]{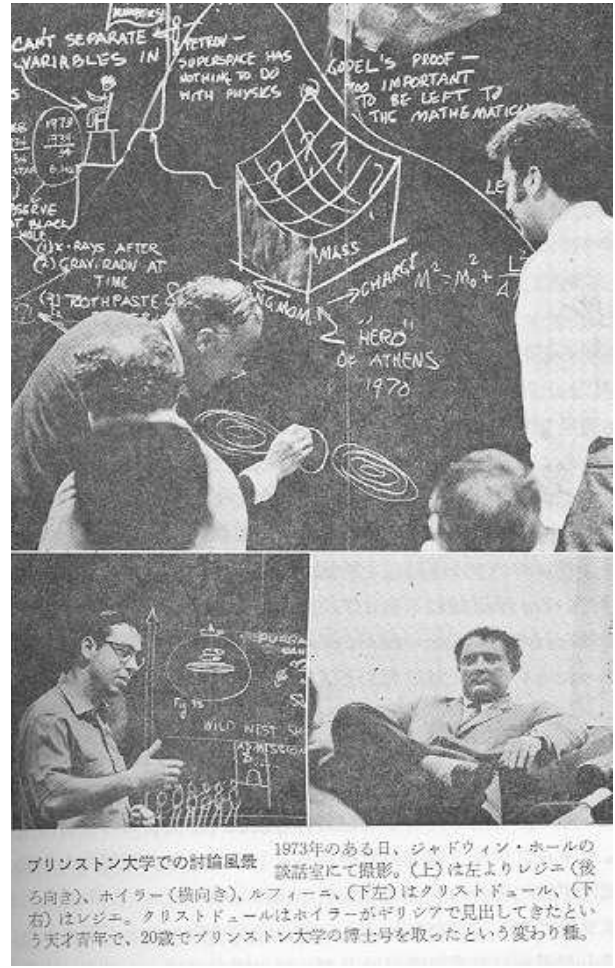}
\end{center}
\caption{Princeton 1971.}
\label{fig1a}
\end{figure}

\begin{figure}[t]
\begin{center}
\includegraphics[width=\hsize,clip]{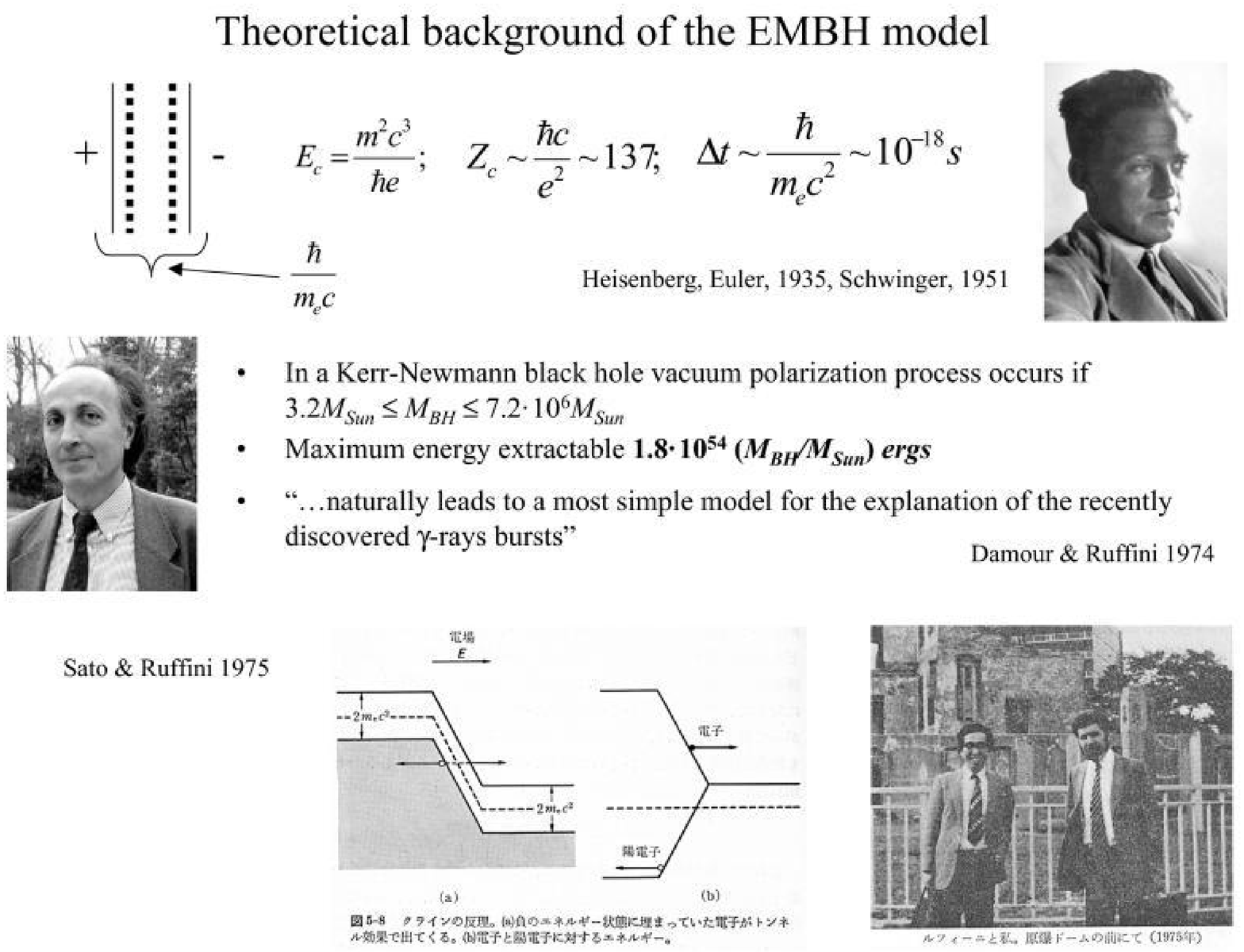}
\end{center}
\caption{The basic components of the Damour \& Ruffini black hole vacuum polarization.}
\label{fig1b}
\end{figure}

My visits to Japan have occurred always in crucial times in the development of our research. In 1975 I visited the University of Kyoto and gave the lectures which were then co-authored in the Japanese book with Humitaka Sato \cite{bookhs}. The focus then was on three major topics:
a) the basic paradigm for the identification of a black hole I had just established and which had found a very significant application in Cygnus X-1 through the splendid data obtained by Riccardo Giacconi and Minoru Oda \cite{bookhs};
b) the Cristodoulou-Ruffini \cite{CR71} mass-energy formula for black holes:
\begin{equation}
E_{BH}^2=M^2c^4=\left(M_{\rm ir}c^2 + \frac{Q^2}{\rho_+}\right)^2 + \frac{L^2c^2}{\rho_+^2}\, ,
\label{em}
\end{equation}
where $M_{\rm ir}$ is the irreducible mass, $\rho_+ = 2(G/c^2) M_{\rm ir}$ is the quasi-spheroidal cylindrical coordinate of the horizon evaluated at the equatorial plane and $Q$ and $L$ are respectively the charge and angular momentum of the black hole; this mass-energy formula allows to estimate the maximum energy extractable from a process of gravitational collapse;
c) a specific energy extraction process from the black hole by pair creation due to supercritical electric fields, first introduced by Sauter \cite{S31}, Heinsenberg \& Euler \cite{HE35}, Schwinger \cite{S51}, I developed with T. Damour \cite{DR75}. In that paper we had also pointed out that such process could be the source of the then newly discovered Gamma-Ray Bursts (GRBs). Our model had a very distinct signature, which differentiates it from all the other models: the characteristic energy of the GRBs should be of the order of $10^{54}$ ergs (see Fig. \ref{fig1a}--\ref{fig1b}).
 
The strategy we had followed, both in the case of Cygnus X1 and the GRBs, was not to try to understand the astrophysical aspects of the phenomenon evidencing the black hole formation. On the contrary we had capitalized on the physics of the black hole and on specific properties of the solutions of Einstein Maxwell Equations in order to infer specific signatures to be expected in the astrophysical scenario in order to obtain the observational evidence for a black hole in a realistic astrophysical setting. Indeed the paradigm for the identification of the black hole in Cygnus X1 (Leach \& Ruffini \cite{lr73}) was based mainly on three general relativistic considerations:
a) the comprehension of the gravitational binding energies around a Kerr black hole, which I found with Wheeler in 1969 \cite{ll}, clearly pointing to the possibility of having accretion energy as the origin of the observed enormous luminosities in X-ray observed in Cygnus X1, $L=10^4L_\odot$;
b) the uniqueness theorem of black hole (see e.g. Ruffini \& Wheeler \cite{rw71}) endowed only of charge mass and angular momentum, clearly pointing to the impossibility of having periodic signals out of a black hole;
c) the existence of an absolute maximum mass of a neutron star, again derived out of first principles, from the equation of equilibrium in the Einstein theory of gravity, the principle of causality implying speed of sound not exceeding the speed of light, and the existence of a fiducial density (Rhoades \& Ruffini \cite{rr74}). All these points were later summarized in the proceedings of the Varenna School organized by Riccardo Giacconi and myself \cite{gr75,gr78} and in the Solvay conference \cite{r74}. Riccardo Giacconi, in his splendid lecture \cite{nobel} recalls the significance of this theoretical work for the understanding of binary X-ray sources.

In the case of GRBs our approach was similar: priority was given to the identification of the energy source of GRBs. That nuclear energy is the energy source of main sequence stars has been credibly proved \cite{sch}, that accretion and gravitational energy release around neutron stars and black holes was the Energy sources of binary x-ray sources had been demonstrated \cite{nobel}, we decided to look in the possibility of having a new energy source as powering the GRBs: the extractable energy of a black hole \cite{CR71,DR75}. The mechanism I had conceived with T. Damour was indeed viable, supported by the very basic and well established physical principles on which it was grounded.

\begin{figure}[t]
\begin{center}
\includegraphics[width=\hsize,clip]{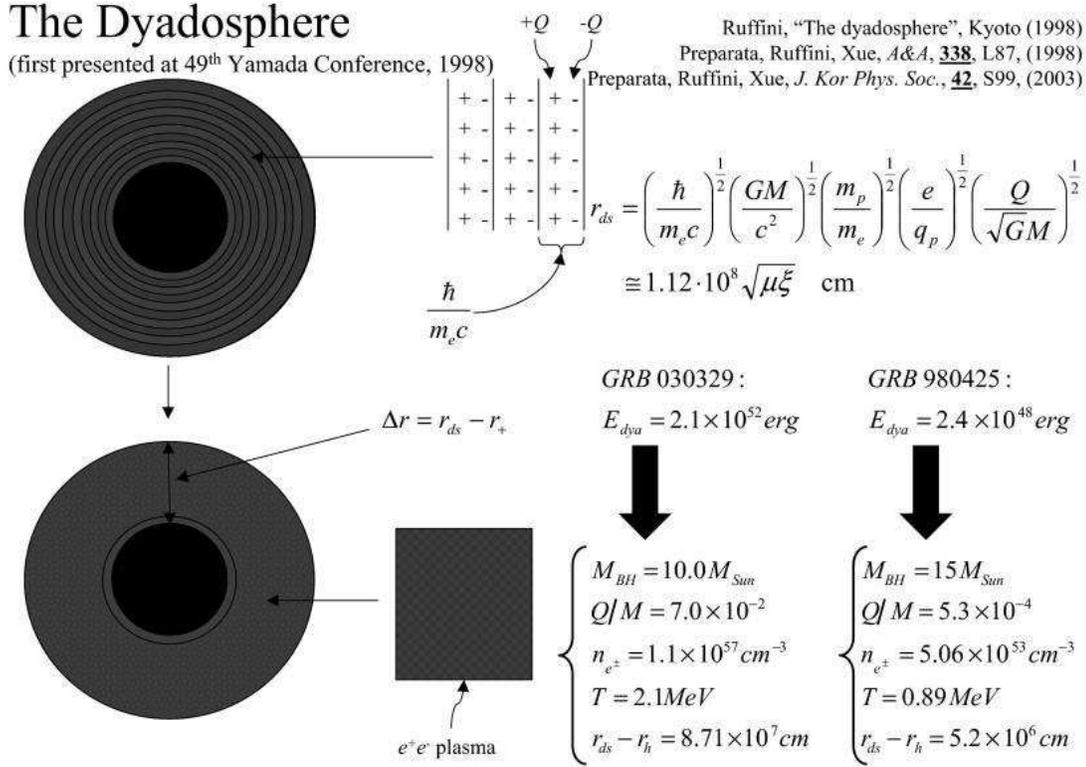}
\end{center}
\caption{The basic parameters of the dyadosphere.}
\label{fig2a}
\end{figure}

\begin{figure}[t]
\begin{center}
\includegraphics[width=8cm,clip]{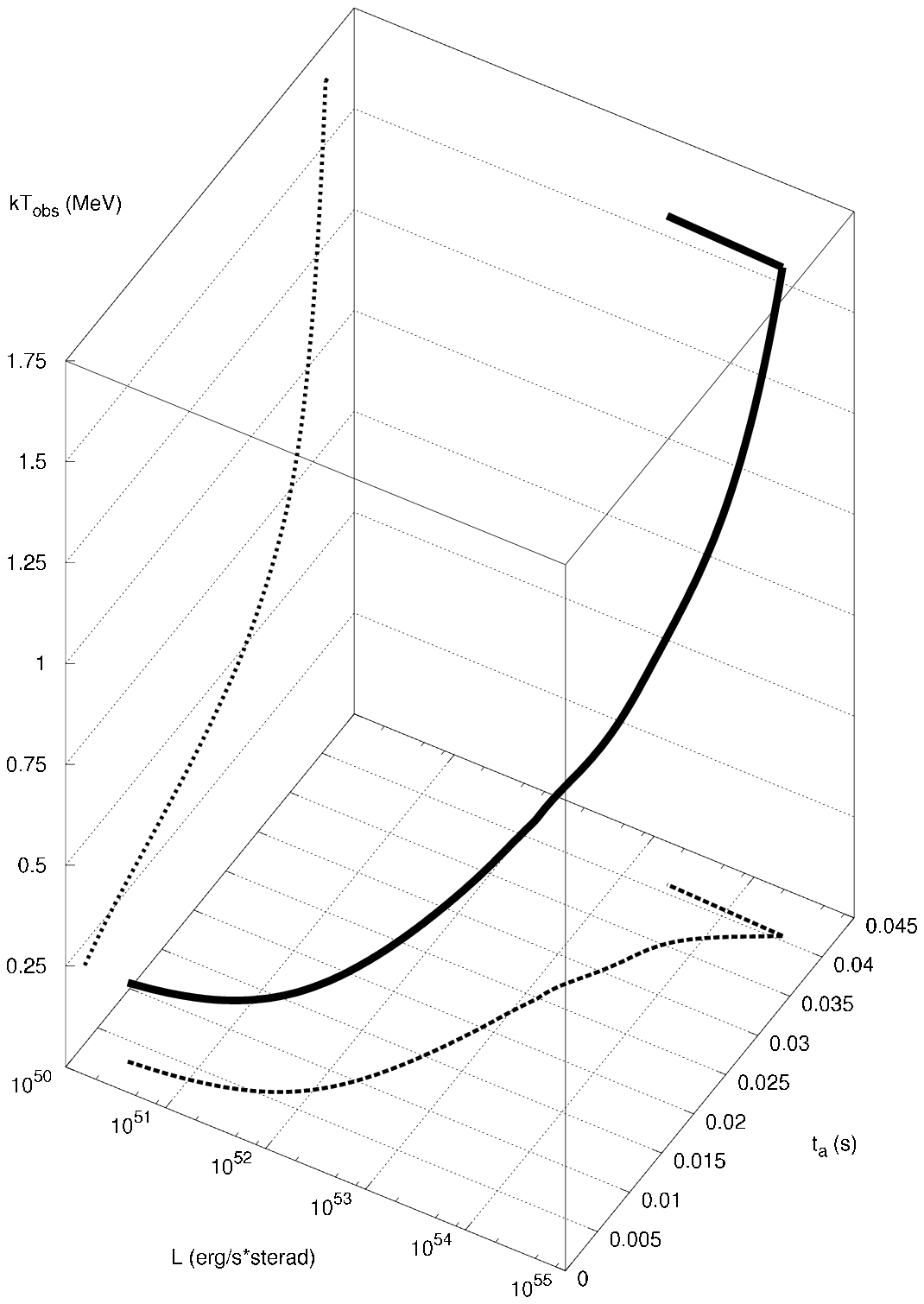}
\end{center}
\caption{The theoretically predicted luminosity and spectral distribution of a short GRB. Details in Ruffini et al. \cite{RFVX03}.}
\label{fig2b}
\end{figure}

My second scientific visit to Japan occurred in occasion of the sixtieth birthday of Humitaka Sato: again I reported \cite{rukyoto} some new progress in our research:
a) The situation with GRBs had dramatically modified by the observations of the Italian-Dutch satellite BeppoSAX (Costa \cite{ca97}) which gave origin to an unprecedented collaboration between X- and $\gamma$-ray, optical and radio astronomy. This observational effort had lead to the determination of the distances of GRBs had unequivocally established the cosmological nature of their source: indeed energetics of the order of $10^{54}$ ergs were implied as predicted by our model with Damour \cite{DR75}.
b) I clarified some basic conceptual issues on the energy extraction process from a black hole endowed with electromagnetic structure, introducing the novel concept of ``dyadosphere'' of a black hole, as the region surrounding the black hole horizon where the electron-positron pairs created in the process of vacuum polarization are localized.
c) I finally pointed out that the very process of thermalization of such an electron-positron plasma created in the dyadosphere is the main mechanism originating the GRB expansion and the engine of the entire GRB phenomenon \cite{RSWX99,RSWX00} (see Fig. \ref{fig2a}--\ref{fig2b}).

In this visit I like to report new results connected with:
a) the physics of the dyadosphere and a possible future verification derived from the observation of short GRBs;
b) the three fundamental paradigms for the theoretical interpretation of GRBs;
c) our understanding of the long bursts and the GRBs afterglows.

\section{On the Dynamical formation of the Dyadosphere and the short GRBs}

The dynamics of the collapse of an electrically-charged stellar core, separating
itself from an oppositely charged remnant in an initially neutral star, was
first modeled by an exact solution of the Einstein-Maxwell equations
corresponding to a shell of charged matter in Ref.~\cite{CRV02}. The
fundamental dynamical equations and their analytic solutions were obtained, 
revealing the amplification of the electromagnetic field strength
during the process of collapse and the asymptotic approach to the final static
configuration. The results, which properly account for general
relativistic effects, are summarized in Fig.~1 and Fig.~2 of Ref.~\cite{CRV02}.
A first step toward the understanding of the process of extracting energy
from a black hole was obtained in Ref.~\cite{RV02}, where it
was shown how the extractable electromagnetic energy is not stored behind the
horizon but is actually distributed all around the black hole. Such a stored
energy is in principle extractable, very efficiently, on time-scales
$\sim\hbar/m_{e}c^{2}$, by a vacuum polarization process \emph{\`{a} l\`{a}}
Sauter-Heisenberg-Euler-Schwinger \cite{S31,HE35,S51}. Such a process occurs
if the electromagnetic field becomes larger than the critical field strength
$\mathcal{E}_{\mathrm{c}}$ for electron-positron pair creation. In
Ref.~\cite{RV02} we followed the approach of Damour and Ruffini \cite{DR75} in
order to evaluate the energy density and the temperature of the created
electron-positron-photon plasma. As a byproduct, a formula for the
irreducible mass of a black hole was also derived solely in terms of the
gravitational, kinetic and rest mass energies of the collapsing core. This
surprising result allowed us in Ref.~\cite{RV03} to obtain a deeper
understanding of the maximum limit for the extractable energy during the process of
gravitational collapse, namely 50\% of the initial energy of
the star: the well known result of a 50\% maximum efficiency for energy extraction in
the case of a Reissner-Nordstr\"{o}m black hole \cite{CR71} then becomes a particular
case of a process of much more general validity.
The crucial issue
of the survival of the electric charge of the collapsing core in the presence of a
copious process of electron-positron pair creation was addressed in
Refs.~\cite{RVX03a,RVX03b}. By using theoretical techniques borrowed from
plasma physics and statistical mechanics
\cite{GKM87,KESCM91,KESCM92,CEKMS93,KME98,SBR...98,BMP...99} based on a
generalized Vlasov equation, it was possible to show that while the core keeps
collapsing, the created electron-positron pairs are entangled in the
overcritical electric field. The electric field itself, due to the back
reaction of the created electron-positron pairs, undergoes damped
oscillations in sign finally settling down to the critical value
$\mathcal{E}_{\mathrm{c}}$. The pairs fully thermalize to an
electron-positron-photon plasma on time-scales typically of the order of
$10^{2}$--$10^{4}\hbar/m_{e}c^{2}$. During this characteristic damping time,
which we recall is much larger than the pair creation time-scale
$\hbar/m_{e}c^{2}$, the core moves
inwards, collapsing with a speed $0.2$--$0.8c$,
further amplifying the electric field strength at its surface and
enhancing the pair creation process.
The first attempt to analyze the
expansion of the newly generated and thermalized electron-positron-photon
plasma was made in Ref.~\cite{RVX03c}. The initial dynamical phases of the
expansion were analyzed, using the general relativistic equations of
Ref.~\cite{CRV02} for the gravitational collapse of the core. The
electron-positron-photon plasma expansion in a sharp pulse of constant
length in the laboratory frame was described following the treatment in
Refs.~\cite{RSWX99,RSWX00}. A {\itshape separatrix} was found in the motion of the plasma
at a critical radius $\bar{R}$: the plasma created at radii larger than
$\bar{R}$ expands to infinity, while the one created at radii smaller than
$\bar{R}$ is trapped by the gravitational field of the collapsing core and
implodes towards the black hole. The value of $\bar{R}$ was found in
Ref.~\cite{RVX03c} to be 
$\bar{R}=2GM/c^{2}[1+\left(  1-3Q^{2}/4GM^{2}\right)^{1/2}]$, 
where $M$ and $Q$ are the mass and the charge of the core, respectively.

In Ruffini et al. \cite{RFVX03} we have described the dynamical phase of the expansion of the pulse of the
optically thick plasma all the way to the point where the transparency
condition is reached. In this process the pulse reaches ultrarelativistic
regimes with Lorentz factor $\gamma\sim10^{2}$--$10^{4}$. The spectra, the
luminosities and the time-sequences of the electromagnetic signals captured
by a far-away observer have been there analyzed in detail for the first time. 
We discretize the gravitational collapse of a spherically symmetric
core of mass $M$ and charge $Q$
by considering a set of events along the world line of a point of fixed
angular position on the collapsing core surface. Between each of
these events we consider a spherical shell slab of plasma of constant coordinate
thickness $\Delta r$.
In order to describe the dynamics of the
expanding plasma pulse the energy-momentum conservation law and the rate equation
for the number of pairs in the Reissner-Nordstr\"{o}m geometry external to
the collapsing core.  The expansion of each slab follows
closely the treatment developed in Refs~\cite{RSWX99,RSWX00} where it was
shown how a homogeneous slab of plasma expands as a pair-electromagnetic
pulse (PEM pulse) of constant thickness in the laboratory frame. The integration stops when each slab
of plasma reaches the optical transparency.

We have integrated the equations for a
core with $M=10M_{\odot},\quad Q=0.1\sqrt{G}M$. We now turn to the results in
Fig.~\ref{fig2} (right), where we plot both the theoretically predicted luminosity $L$
and the spectral hardness of the signal reaching a far-away observer as
functions of the arrival time $t_{a}$. Since all three of these quantities
depend in an essential way on the cosmological redshift factor $z$, see
Refs.~\cite{BRX01,RBCFX03}, we have adopted a cosmological
redshift $z=1$ for this figure.

The projection of the plot in Fig.~\ref{fig2} (right) onto the $k\gamma T_{\mathrm{obs}%
}$, $t_{a}$ plane describes the temporal evolution of the spectral hardness.
We observe a precise soft-to-hard evolution of the spectrum of the gamma ray signal from $\sim10^{2}$ keV monotonically increasing to $\sim1$ MeV. We
recall that $kT_{\mathrm{obs}}=k\gamma T/\left(  1+z\right)$. The
above quantities are clearly functions of the cosmological redshift $z$, of
the charge $Q$ and the mass $M$ of the collapsing core.

The arrival time interval is very
sensitive to the mass of the black hole: $\Delta t_{a}\sim10^{-2}-10^{-1}$ s. Similarly the spectral hardness of the signal is sensitive to the ratio
$Q/\sqrt{G}M$ \cite{RFVX04}. Moreover the duration, the spectral hardness and
luminosity are all sensitive to the cosmological redshift $z$ (see
Ref.~\cite{RFVX04}).
The characteristic spectra, time variabilities and luminosities of the
electromagnetic signals from collapsing overcritical stellar cores we have derived agrees closely with the observations of
short-GRBs \cite{P...99}. New space missions should possibly be planned, with temporal resolution down
to fractions of $\mu$s and higher collecting area and spectral resolution,
in order to verify the agreement between our model
and the observations. It is now clear that if our theoretical
predictions will be confirmed, we would have a very powerful tool for
cosmological observations: the independent information about luminosity,
time-scale and spectrum can uniquely determine the mass, the electromagnetic
structure and the distance from the observer of the collapsing core, see Ref.~\cite{RFVX04}.
In that case short-GRBs may become the best example of standard candles in cosmology \cite{rbbcflx04}.

\section{The three paradigms for the interpretation of GRBs}

I recall briefly the three paradigms which we have introduced in 2001 as a guideline for the understanding of GRBs:\\
{\bf 1)} The Relative Space-Time Transformation (RSTT) paradigm \cite{lett1}: ``the necessary condition in order to interpret the GRB data, given in terms of the arrival time at the detector, is the knowledge of the {\em entire} worldline of the source from the gravitational collapse. In order to meet this condition, given a proper theoretical description and the correct constitutive equations, it is sufficient to know the energy of the dyadosphere and the mass of the remnant of the progenitor star''.\\
{\bf 2)} The Interpretation of the Burst Structure (IBS) paradigm \cite{lett2}. In it we reconsider the relative roles of the afterglow and the burst in the GRBs by defining in this complex phenomenon two new phases: {\bf a)} the {\em injector phase} starting with the process of gravitational collapse, encompassing the above Eras I, II, III and ending with the emission of the Proper-GRB (P-GRB); {\bf b)} the {\em beam-target phase} encompassing the above Eras IV and V giving rise to the afterglow. In particular in the afterglow three different regimes are present for the average bolometric intensity : one increasing with arrival time, a second one with an Extended Afterglow Peak Emission (E-APE) and finally one decreasing as a function of the arrival time.\\
{\bf 3)} The GRB-supernova time sequence (GSTS) paradigm introduces the concept of {\em induced supernova explosion} in the supernovae-GRB association \cite{lett3} leading to the very novel possibility of a process of gravitational collapse induced on a companion star in a very special evolution phase by the GRB explosion.

\section{Selected examples of the understanding of GRBs}

As a test of the understanding of GRBs, I illustrated in the oral presentation the case of GRB991216, GRB980425 and GRB030329 which will not be reported in this printed version since they are contained in the published literature and I am going to give uniquely the list of references. GRB991216 (Ruffini et al. \cite{lett1,lett2,lett3,lett5,LongArticle,cnr02,rbcfx02,Brasile}) has been used as a prototype in order to prove the validity of the first two paradigms, while GRB980425 (Ruffini et al. \cite{cnr02,cospar02,r03mg10,f03mg10,f03la}) and GRB030329 (Ruffini et al. \cite{r03mg10,b03mg10,b03la}) have been a test of all the above three paradigms (see Figs. \ref{980425a}--\ref{030329b}).

\begin{figure}[t]
\begin{center}
\includegraphics[width=\hsize,clip]{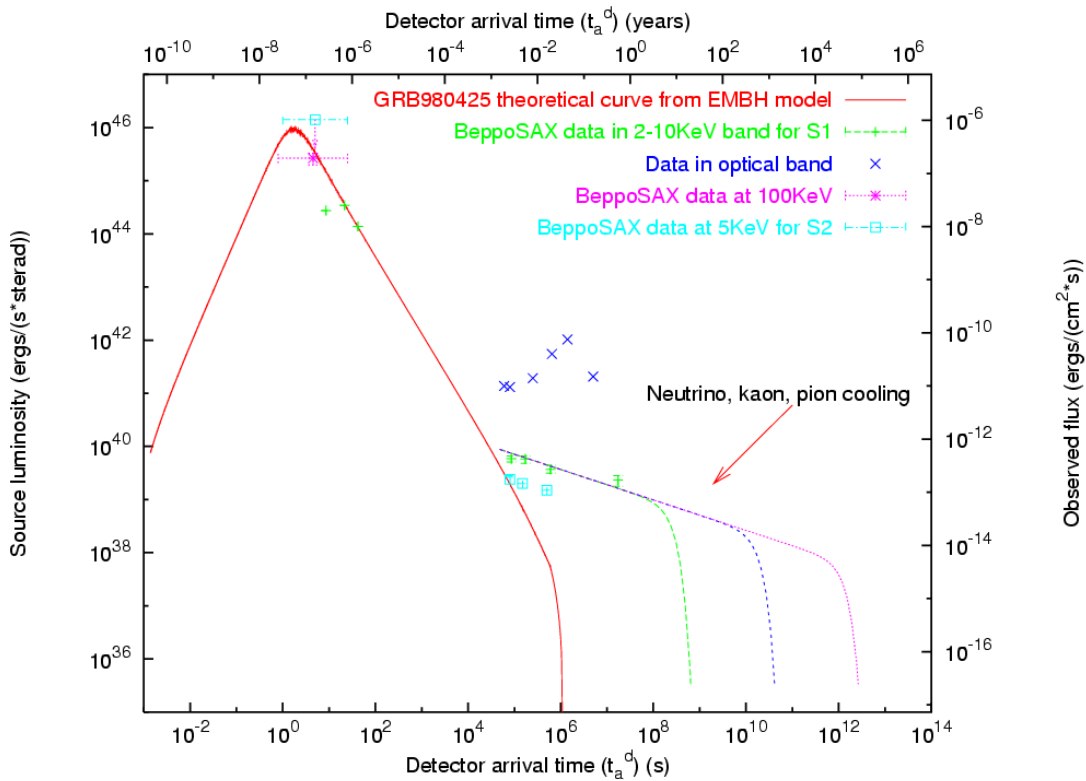}
\end{center}
\caption{Theoretical fit of the $\gamma$ and X-ray luminosity in the afterglow of GRB 980425. Details in Ruffini et al. \cite{cnr02,cospar02,r03mg10,f03mg10,f03la}}
\label{980425}
\end{figure}

\begin{figure}[t]
\begin{center}
\includegraphics[width=\hsize,clip]{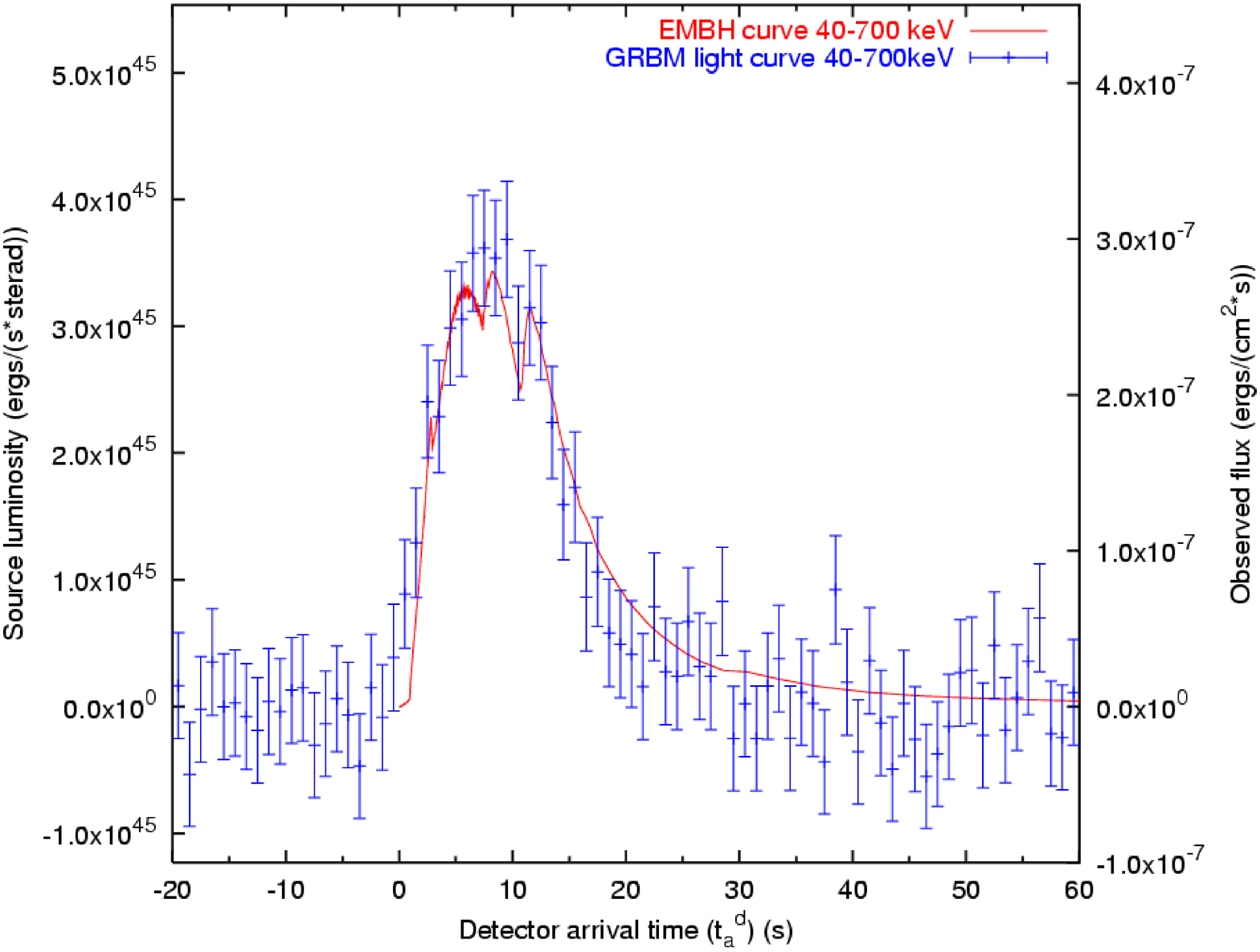}
\end{center}
\caption{Theoretical fit of the $\gamma$ and X-ray luminosity in the E-APE of GRB 980425. Details in Ruffini et al. \cite{cnr02,cospar02,r03mg10,f03mg10,f03la}}
\label{980425b}
\end{figure}

\begin{figure}[t]
\begin{center}
\includegraphics[width=\hsize,clip]{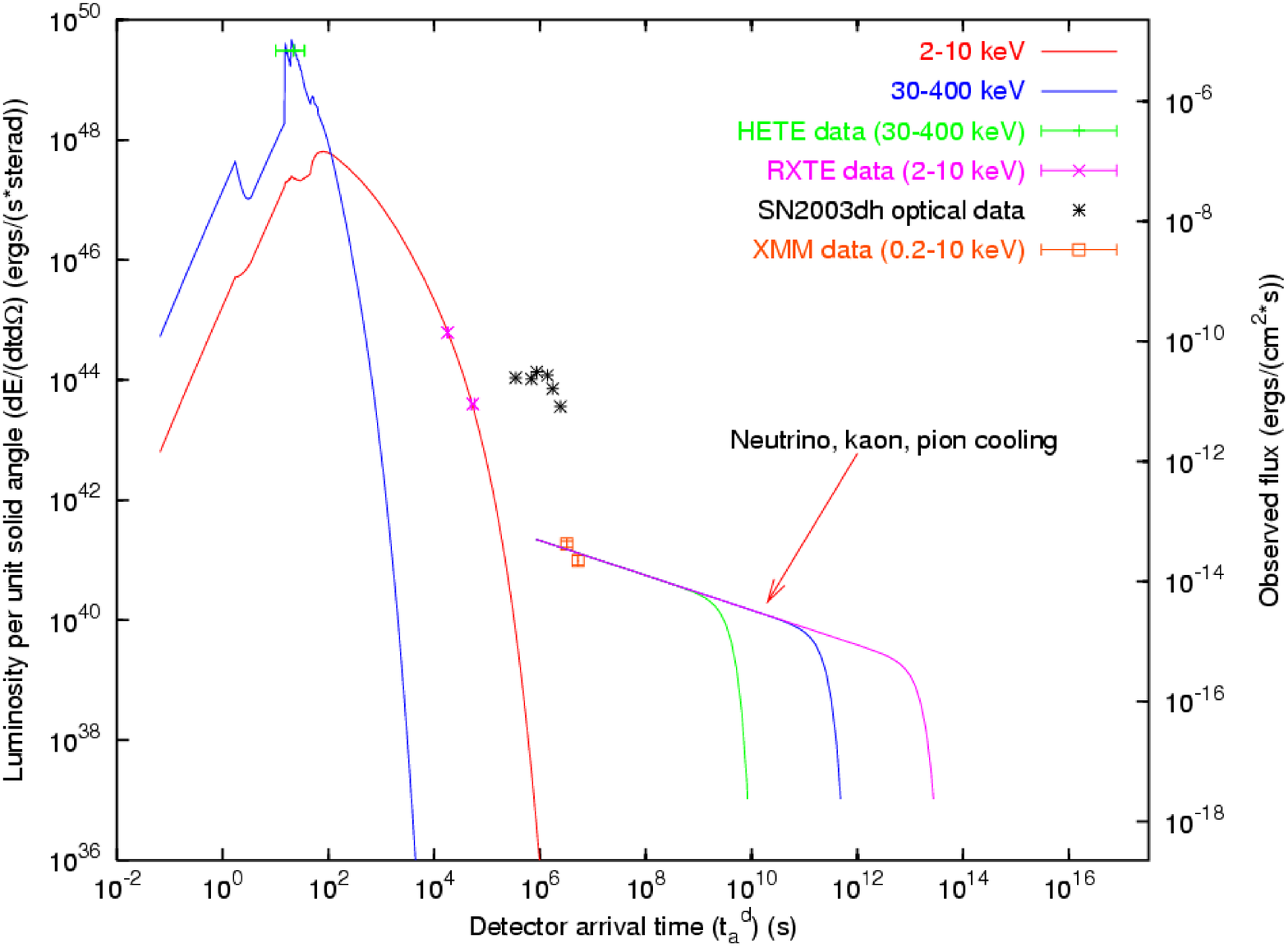}
\end{center}
\caption{Theoretical fit of the $\gamma$ and X-ray luminosity in the afterglow of GRB 030329. Details in Ruffini et al. \cite{r03mg10,b03mg10,b03la}}.
\label{030329a}
\end{figure}

\begin{figure}[t]
\begin{center}
\includegraphics[width=\hsize,clip]{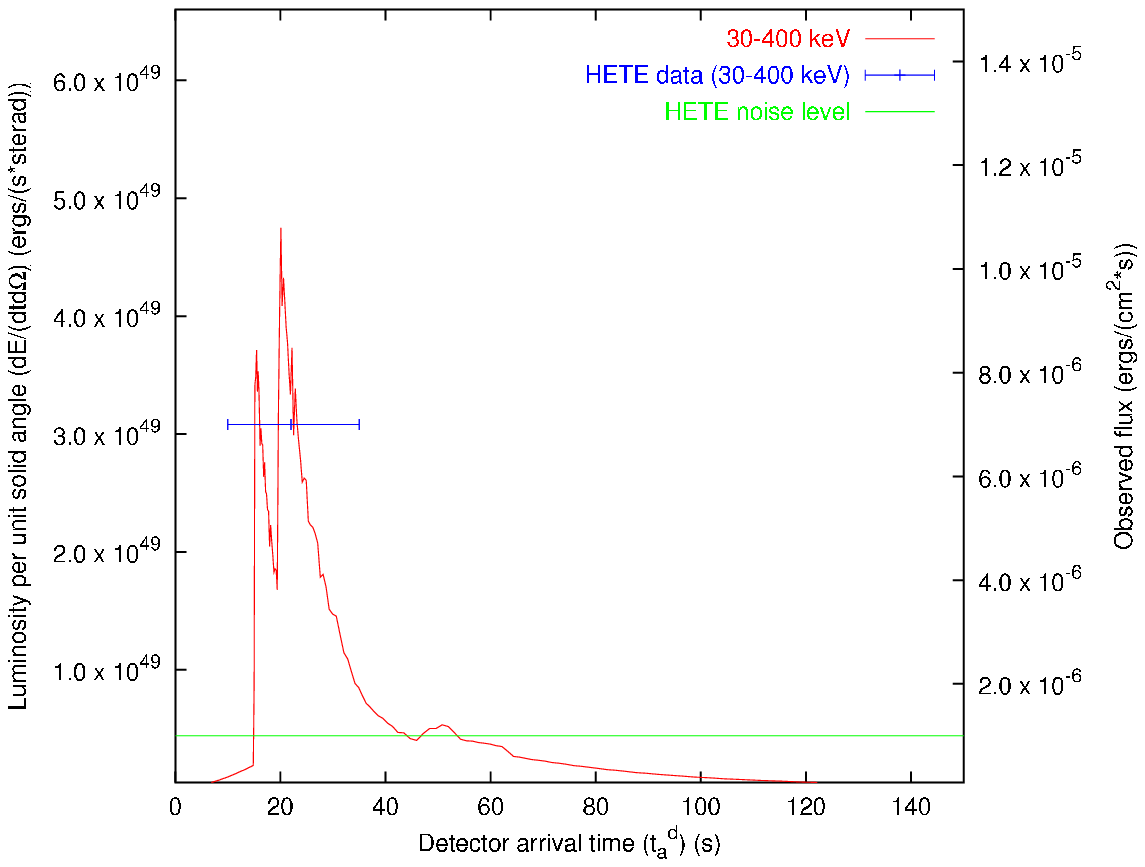}
\end{center}
\caption{Theoretical fit of the $\gamma$ and X-ray luminosity in the E-APE of GRB 030329. Details in Ruffini et al. \cite{r03mg10,b03mg10,b03la}}.
\label{030329b}
\end{figure}

\end{document}